\documentstyle[prl,aps,multicol,epsf]{revtex}
\input psfig
\def \be{\begin{equation}}
\def \ee{\end{equation}}

\def \o{\omega}

\def \e{\epsilon}

\def \H{{\cal H}}

\def \R{G^{(R)}}

\begin{document}

\bibliographystyle{simpl1}

\title{Theory of Dephasing by External Perturbation in Open 
Quantum Dots}
\date{\today}
\author{Maxim G. Vavilov$^a$ and Igor L. Aleiner$^b$}
\address{
$^a$ Laboratory of Atomic and Solid State Physics,
Cornell University, Ithaca NY 14853\\
$^b$ Department of Physics and Astronomy, SUNY at Stony Brook,
Stony Brook, NY 11794
}
\maketitle
\begin{abstract}
We propose a random matrix theory describing the influence of a time dependent 
external field on the average magnetoresistance of open quantum
dots. The effect is taken into account in all orders of perturbation theory, 
and the result is applicable to both weak and strong external
fields.
\end{abstract}
\draft
\pacs{PACS numbers: 73.23.Ad, 72.15.Rn, 72.70.+m}
\begin{multicols}{2}

It is well established that the anomalous magnetoresistance of 
bulk disordered systems is governed 
by the weak localization (WL) \cite{Altshuler80,AA,AAKL}. Being
an interference phenomenon, the WL is extremely
sensitive to inelastic processes which are commonly referred to as
dephasing. 

Recently, another object for the studying the quantum effects appeared
-- ballistic quantum dots\cite{Review}. In the absence of 
inelastic processes, the transport properties of the dots are well
described within the Random Matrix Theory (RMT)\cite{Baranger1}. The
magnetoresistance within this theory manifests itself as a crossover
between two universal ensembles (orthogonal and unitary), and the
strength of the magnetic field defines a position on that crossover.
This approach per se does not include dephasing, and the dephasing processes
were considered on a phenomenological basis \cite{BM1}.
The relation of this phenomenological description 
used to fit the data of Ref.\cite{dephexp}
to microscopic mechanisms of dephasing is still an open question.

In this Letter, we propose a Random Matrix - like theory 
of the magnetoresistance affected by an external time dependent
perturbation. We will be able to find both the amplitude and
frequency dependence of the magnetoresistance using only one unknown
parameter. This parameter can be related to the correlator of the
level velocities due to the same perturbation at zero  frequency and,
thus, in principle can be determined by an independent experiment. 
After the
strength of the potential is normalized by this parameter, all the
results become universal. From the experimental point of view, we have
in mind changing shape of a quantum dot by applying an external {\em ac}- bias.

Before we proceed, let us mention that the effect studied in the
present paper is similar to that of Ref.~\cite{AAK}, where it was shown
that uniform ac- electric filed suppresses the weak localization
correction to the conductivity of a disordered wire (experimentally
it was studied in Ref.~\cite{Vea}).  However the results of \cite{AAK}
are not directly applicable to the quantum dots with size $L$ so small
that the Thouless energy $E_T\sim \hbar/\tau_{erg}$ is much greater than
other energy scales (such as the dephasing or escape rates) of the problem
(here $\tau_{erg}$ is the characteristic time for the classical
particle to cover all of the available phase space).  
  
On the other hand in
this limit  one can use the RMT to 
study the conductance of the system, see \cite{B}. All 
corrections to the RMT are small as
$N_{ch}/g_{dot}$; $g_{dot}=E_T/\delta_1$ and $\delta_1$ is the mean
level spacing.
We consider the WL correction to the conductance of quantum 
dots with the large number $N_{ch}$ of open 
 channels. In this approximation we neglect the effect 
of interaction on the conductance which are small 
as $1/N_{ch}^2$\cite{Brouwer}, while the 
weak localization correction to the conductance is proportional
to $1/N_{ch}$. The same condition also  allows us to use
conventional diagrammatic technique\cite{AGD} to 
perform the ensemble average. 
 
In general, the Hamiltonian of the system can be represented as, see 
\cite{B}:
\be
\label{1}
\hat H=\hat H_D+\hat H_L+\hat H_{LD},
\ee 
where $\hat H_D$ is the Hamiltonian of electrons in the dot,
which is determined by $M\times M$ matrix
\be
\label{2}
\hat H_D=\sum_{n,m=1}^M \psi^\dag_n H_{nm} \psi_m,
\ee
where the thermodynamic limit $M \to\infty$ is assumed.
 We consider the case, when $H_{nm}$ is a time  dependent random matrix
in the form:
\be
\label{3}
H_{nm}(t)=\H_{nm}+V_{nm} \varphi(t).
\ee
Here the time independent part of the Hamiltonian $\H_{nm}$ is a
random realization of $M\times M$ matrix, which obeys the correlation
function 
\be
\label{4}
\langle\H_{nm}\H^*_{n'm'}\rangle=\lambda\delta_{nn'}\delta_{mm'}
+\lambda'\delta_{mn'}\delta_{nm'}, \ee where
$\lambda=M(\delta_1/\pi)^2$ and $\lambda'=\lambda(1-g_h/4M)$, 
$g_h$ defines crossover from
the orthogonal ($g_h=0$) to the unitary ($g_h=4M$) ensemble.  
Parameter $g_h$
has a meaning of the dephasing rate by the external magnetic field in
units of the level spacing $\delta_1$, see \cite{AAKL,B}.
It can be estimated as $g_h \simeq g_{dot} \left(\Phi/\Phi_0\right)^2$
where $\Phi$ is the magnetic flux through the dot and $\Phi_0=hc/e$ is
the flux quantum. The time dependent perturbation is described by
symmetric $M\times M$ matrices $V_{nm}$ and a function of time $\varphi(t)$.

The coupling between the dot and the leads is
\be
\label{5}
\hat H_{LD}=\sum_{\alpha, n, k}\left( W_{n \alpha}\psi^\dag_\alpha
(k)
\psi_n+
{\rm H.c.}\right),
\ee
where $\psi_n$ correspond to the states of the dot,
$\psi_{\alpha}(k)$ denotes different electron states in the leads,
and momentum $k$ labels continuous state in each channel $\alpha$.
For the dot connected with two leads by $N_l$ and
$N_r$ channels respectively,
we denote the left lead channels by 
$1\leq \alpha\leq N_l$ and the right channels by 
$N_l+1\leq \alpha \leq N_{ch}$, where 
$N_{ch}=N_l+N_r$.
The  electron spectrum in the leads near Fermi
surface can be linearized:
\be
\label{6}
\hat H_L=v_F\sum_{\alpha, k} k \psi^\dag_\alpha(k)\psi_\alpha(k),
\ee
where  $v_F=1/2\pi \nu$ is the
Fermi velocity and $\nu$ is the density of states at the Fermi surface.

The coupling constants $W_{n \alpha}$ in Eq.(\ref{5})are defined as\cite{B}:
\be
\label{7}
W_{n \alpha}=\sqrt{\frac{M\delta_1}{\pi^2\nu}}\cases{
t_\alpha,& if $n=\alpha\leq N_{ch}$,\cr`
0,& otherwise,
}
\ee
where $t_\alpha$ determines the dimensionless
conductance of each lead (in units of $2e^2/h$) according to
\be
\label{7.5}
\begin{array}{ccc}
\displaystyle
g_l=\sum_{\alpha=1}^{N_l} 
\frac{4t_\alpha t^*_\alpha}{(1+t_\alpha t^*_\alpha)^2},
& &
\displaystyle
g_r=\sum_{\alpha=N_l+1}^{N_{ch}} 
\frac{4t_\alpha t^*_\alpha}{(1+t_\alpha t^*_\alpha)^2}
\end{array}
\ee
and $|t_\alpha|\leq 1$.
The factor in Eq.~(\ref{7}) is chosen so that
the ensemble average scattering matrix ${\cal S}_{\alpha\beta}$ of a
dot with fully open channels
($t_\alpha=1$) is zero. More complicated structure of $\hat W$ can be
always reduced to the form (\ref{7}) by suitable rotations. 

For the system described above the scattering matrix 
$\hat {\cal S}$ has the form:
\be
\label{8}
{\cal S}_{\alpha\beta}(t,t')=1-2\pi i\nu 
W^\dag_{\alpha n} G_{nm}(t,t') W_{m\beta},
\ee 
and the Green's function $G_{nm}(t,t')$ is the solution to:
\be
\label{9}
\left(i\frac{\partial}{\partial t}-{\hat H}(t)+
i\pi\nu \hat{W}\hat{W}^\dag \right)\hat{G}(t,t')=
\delta(t-t'),
\ee
where matrices $\hat{H}$ and $\hat{W}$ are
 comprised by their elements (\ref{3}) and (\ref{7}) respectively.
 
The averaged 
dimensionless $dc$ - conductance of the dot is determined in terms of the 
scattering matrix of the system in the linear response theory by 
(see, e.g. \cite{B}):
\be
\label{10}
 g ={\Big \langle} \int_{-\infty}^t dt'  {\mathrm tr}
\left[\hat \tau_l {\cal S}(t,t') \hat
\tau_r {\cal S}^\dag (t,t')\right]{\Big \rangle},
\ee
where $\langle\dots\rangle$ stands for both ensemble and time averages.
We also introduced notation for the projector on the left lead,
 $\hat \tau_{l}$, which is a diagonal $N_{ch} \times N_{ch}$
matrix with the
first $N_l$ diagonal elements equal to unity, 
and the other diagonal
elements equal to zero, and $\hat \tau_r=\hat{I}-\hat \tau_{l}$.

We perform calculations of the average conductance keeping the leading 
terms in $1/M$.
The diagrammatic technique is somewhat similar to that developed for 
bulk metals \cite{AGD}, where the small parameter is
$1/\e_F\tau_{imp}$ with $\e_F$ being the Fermi energy and $\tau_{imp}$
being the elastic mean free time.

\begin{figure}
\centerline{\psfig{figure=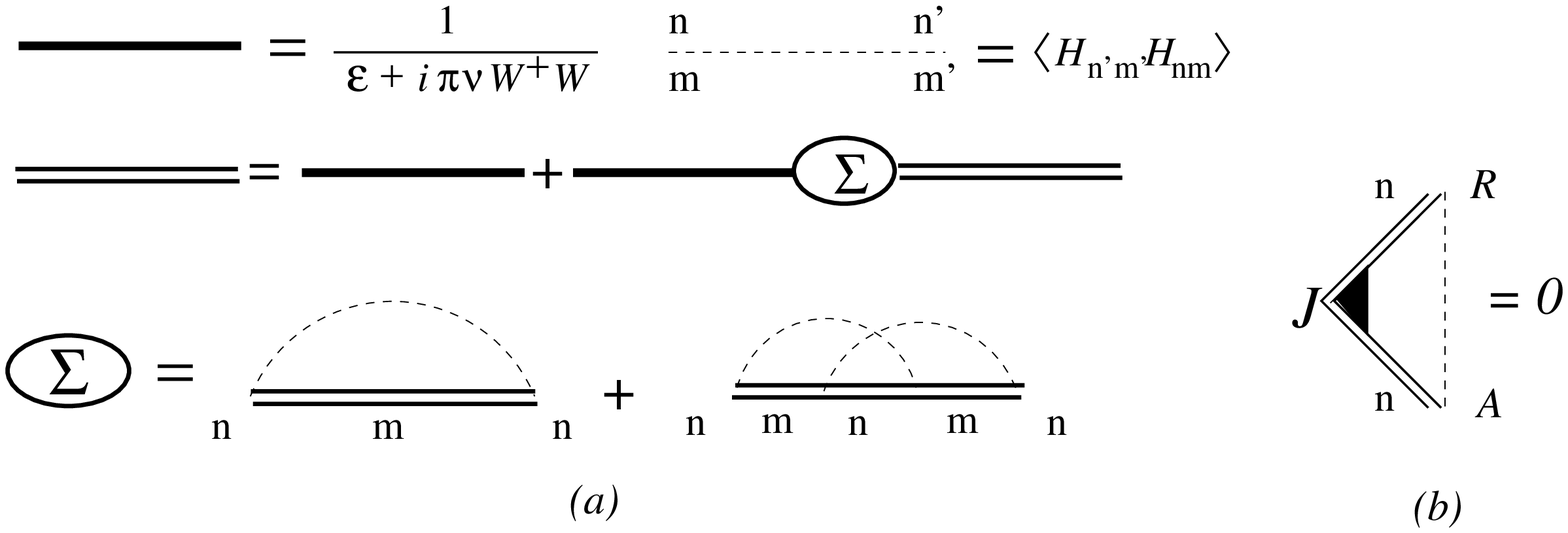,width=8.5cm}}
\narrowtext
\caption{
(a) Diagrams for the ensemble average Green's function.  
The second term in the self-energy
includes an intersection of dashed lines and it is 
small as $1/M$. (b) The representation of the conductance in the form
of Eq.~(\ref{12})  forbids
the renormalization of vertices $J$ from Eq.~(\ref{12.5}) by disorder.}
\end{figure}

First let us find the ensemble average Green's function 
$\langle \R\rangle$. 
One can see  that $\langle \R\rangle$ is diagonal, 
$\langle\R_{nm}(\e)\rangle = \delta_{nm}\R_{n}(\e)$ 
Using the self consistency equation for the Green's function, Fig.~1
(a), we find
\be
\label{11}
\R_{n}(\e)=\frac{1}{i\sqrt{\lambda M}}
\cases{
\displaystyle
\frac{1}{1+t_nt_n^*}, & $n\leq N_{ch}$, \cr
\displaystyle
1+\frac{\sum_{\alpha=1}^{N_{ch}}
\frac{2t_\alpha t_\alpha^*}{1+t_\alpha t_\alpha^*}+i\e}{4M}, 
& $n>N_{ch}$.
}
\ee
Here we introduced the dimensionless energy $\e$ measured in units of 
$\sqrt{\lambda/4M}=\delta_1/2\pi$. We expand these Green's 
functions in $\e/M$ and $(g_l+g_r)/M$, since only those terms survive 
the thermodynamic limit
$M\to\infty$. For the same reason, the expression for $\R_{n}$ with 
$n\leq N_{ch}$  
neglects such terms at all because the contribution of those elements
to the final answer is already small as $N_{ch}/M$.

To simplify further manipulations,
we rearrange Eq.~(\ref{10}) in the following form
\begin{eqnarray}
\label{12}
\displaystyle
g(t)&=&{\Big\langle}\int\limits_{-\infty}^{t}dt'
{\rm tr}\left[ \hat J_l{\cal S}(t,t')
\hat J_r{\cal S}^\dag (t,t')\right]{\Big\rangle}\\
\displaystyle
&+&\frac{N_r g_l^2 +N_lg_r^2}{(g_l+g_r)^2},
\nonumber
\\
\label{12.5}
\hat J_{l,r}&=&\hat \tau_{l,r}-\frac{g_{l,r}}{g_l+g_r}\hat{I}.
\end{eqnarray}
Equation~(\ref{12}) immediately follows from Eq.~(\ref{10}) and the unitarity
of the ${\cal S}$- matrix ${\cal S}{\cal S}^\dag=1$. 
The calculations of the conductance in the form of Eq.~(\ref{12}) 
are significantly simpler since
the vertices (\ref{12.5}) are not dressed by dashed lines, see Fig.~1(b).
This trick is similar to the calculation of the conductivity of disordered
bulk systems in terms of the current-current
instead of density-density correlation function, see Refs.~\cite{AA,AAKL}. 

Now we substitute the scattering matrix defined by Eq.~(\ref{8}) to
Eq.~(\ref{12}). To the leading order in $1/(g_l+g_r)$ one can average
${\cal S}$ -matrices independently with the help of Eq.~(\ref{11}) and 
obtain the classical conductance 
\be
\label{14}
 g_{cl} = \frac{g_lg_r}{g_l+g_r}.
\ee 
In particular, for the dot with fully open channels ($t_\alpha=1$), the
averaged ${\cal S}$- matrix vanishes ($\langle{\cal S}\rangle=0$) 
and the last term of 
Eq.~(\ref{12}) gives the known result $  g_{cl} 
=N_lN_r/N_{ch}$, since in this case $g_{l,r}=N_{l,r}$.

\begin{figure}
\centerline{\psfig{figure=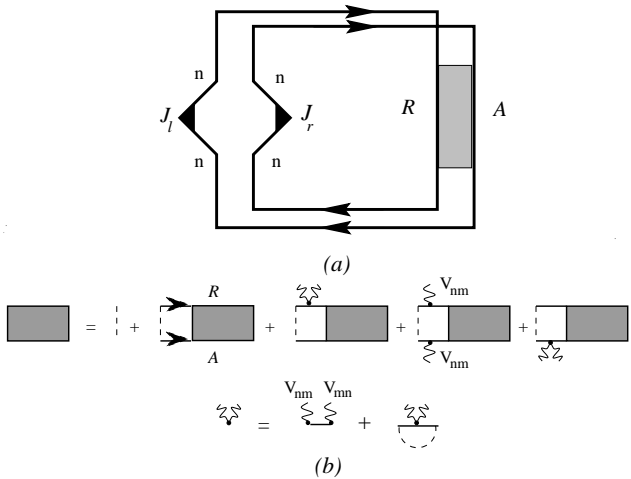,width=8.5cm}}
\narrowtext
\caption{
(a) The diagram for the WL correction to
the conductance.
(b) The diagram equation for the Cooperon.
}
\end{figure}

The first order correction in 
$1/(g_l+g_r)$ to Eq.~(\ref{14}) is given by the
diagram in Fig.~2(a). 
It represents the WL correction to the 
conductance, and has the analytic expression
\be
\label{19}
 \Delta g_{wl}=-
\frac{f_lg_r^2+f_rg^2_l}{(g_l+g_r)^2}
\int\limits_0^{2\pi/\o}\! \frac{ \omega d T}{2\pi}
\int\limits_{0}^\infty\! 2d\tau 
{\cal C}(T,\tau,-\tau)
, 
\ee
where formfactors $f_{l,r}$ are given by
\be
\label{20}
\begin{array}{lr}
\displaystyle
f_l=\sum_{\alpha=1}^{N_l}\frac{16(t_\alpha t^*_\alpha)^2}
{(1+t_\alpha t^*_\alpha)^4},&
\displaystyle
f_r=\sum_{\alpha=N_l+1}^{N_{ch}}\frac{16(t_\alpha t^*_\alpha)^2}
{(1+t_\alpha t^*_\alpha)^4}.
\end{array}
\ee

The Cooperon ${\cal C}$ is defined by  Fig.~2(b): 
\be
\label{15}
\left(\frac{\partial}{\partial \tau}+{\cal K}(T,\tau)\right)
{\cal C}(T,\tau,\tau')=
\delta(\tau-\tau'), 
\ee
where time is measured in units of inverse level spacing $2\pi/\delta_1$
and the ``Hamiltonian'' for the Cooperon is 
\be
\label{16}
{\cal K}(T,\tau)=g_*+ \pi^2 C_0
\left[\varphi(T+\tau/2)-\varphi(T-\tau/2)\right]^2, 
\ee 
with $g_*$
characterizes the total dephasing due to the escape and the magnetic
field, $g_*=g_l+g_r+g_h$, and we chose $\varphi(t)=\cos \o t$ 
to describe the time dependence of the perturbation.

The only unknown parameter, $C_0$, in Eq.~(\ref{16}) depends on the strength
of the perturbation. In terms of the original 
Hamiltonian (\ref{3}), it is defined as
\be
\label{17}
{C}_0=\frac{2}{\pi^2 M\lambda} \sum_{nm}V^2_{nm},  
\ee
where we used the fact that the matrix $\hat{V}$ is symmetric.  This
parameter is also related to the typical value of the level
velocities, which characterizes the evolution of energy levels 
$\epsilon_{\nu}(X)$ under 
the action of the external perturbation $X\hat{V}$, see \cite{E}:
\be
\label{18}
\delta_1^2 C_0=
{\Big\langle}
\left(\frac{\partial \e_\nu}{\partial X}\right)^2
{\Big\rangle}
-
{\Big\langle}
\frac{\partial \e_\nu}{\partial X}
{\Big\rangle}^2.
\ee
Since all other responses (e.g. parametric dependence of the
conductance of the dot) are expressed in terms of 
universal functions of the same parameter $C_0$ \cite{E}, it can
be found from independent measurements. For not very realistic case
 of homogeneous electric field $E$ introduced into the dot of linear
size $L$, one can estimate $C_0 \simeq (eEL)^2/(E_T\delta_1)$.
It is important to emphasize that the homogeneous shift of all 
levels does not affect the magnetoresistance and that is why the
average level velocity $\langle\partial \e_\nu/\partial X\rangle$ is not relevant.

In the absence of the time dependent perturbation $\varphi \equiv 0$,
one obtains 
\cite{B,BB96} from Eqs.~(\ref{19})--(\ref{16}):
\be
\label{190}
 \Delta g_{wl}^{(0)}=-
\frac{f_lg_r^2+f_rg^2_l}{(g_l+g_r)^2g_*}.
\ee
The solution to Eq.~(\ref{15}) gives the weak localization correction
to the conductance 
$ \Delta g_{wl}$
in the presence of the time dependent field. It can be 
expressed in terms of the unperturbed correction (\ref{190}) as  
\be
\label{21}
\frac{ \Delta g_{wl} }
{ \Delta g^{(0)}_{wl} }=
F\left(y,z\right),
\quad y = \frac{\pi \omega}{g_*\delta_1},\ z= \frac{\pi^2 C_0}{g_*},
\ee
where dimensionless function $F(y,z)$ is given by
\be
\label{22}
F(y,z)=\!
\int_0^\infty\! d x e^{-x-z\phi}
{\mathrm I}_0\left[z\phi\right], \quad \phi = x- \frac{\sin xy}{y}. 
\ee
Here ${\mathrm I}_0(\xi)$ is the modified Bessel function.
Some curves for this function are plotted in Fig.~3.

Equations (\ref{21}) -- (\ref{22}) are the main results
of our paper. They give the universal description of the effect of the 
external field on the weak localization correction. Below
we will discuss different asymptotic regimes and compare
them with the results for bulk systems \cite{AAKL}.

For weak external field $z \ll {\mathrm max} (1,y^{-2})$
we find
\be
\label{24}
\frac{ \Delta g_{wl} }
{ \Delta g^{(0)}_{wl} }=
1-\frac{\pi^2C_0}{g_*}\frac{\pi^2\o^2}
{\pi^2\o^2+\delta_1^2 g_*^2}.
\ee 
In this regime the correction is quadratic in the frequency for
slowly oscillating field, similarly to the bulk system result at $\omega$
smaller than the dephasing rate $1/\tau_\phi$. However, 
the frequency dependence saturates at large frequency. 
It is different from the result for bulk systems, where
a characteristic spatial scale shrinks
as $1/\sqrt{\omega}$, whereas in our case it is determined by 
the size of the dot. 
 
\begin{figure}
\centerline{\psfig{figure=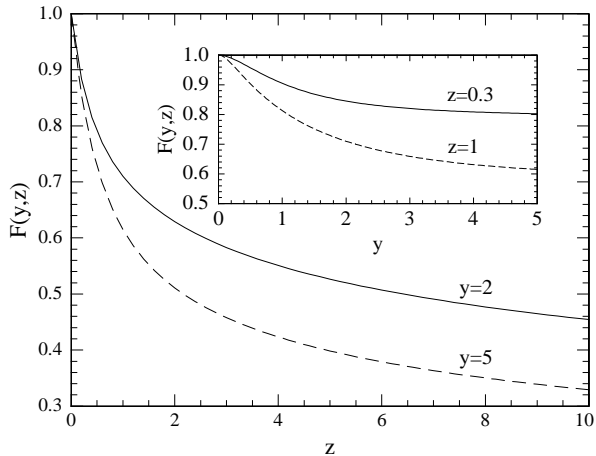,width=8.5cm}}
\narrowtext
\caption{
Representative curves of $F(y,z)$ as a function of $z$ 
for two values of $y$. It decreases linearly with $z$ at small 
values of $z$.
The inset shows the $y-$dependence of the function $F(y,z)$ 
for two values of $z$.
It decreases quadratically in $y$ at small values of 
$y$ and saturates at larger $y$.
}
\end{figure}

In the opposite limit of strong external field 
$z \gg {\mathrm max} (1,y^{-2})$ we have to consider separately the 
cases of fast, $y \gg 1$, and slow, $y \ll 1$ field oscillations.
In the first case we find 
\be
\label{25}
\frac{ \Delta g_{wl} }
{ \Delta g^{(0)}_{wl} }=\sqrt{\frac{g_*}{2\pi^2 C_0}}.
\ee
The linear dependence of the quantum correction on 
$1/\sqrt{C_0}$ is similar to that for the bulk system.
Contrary to the bulk systems, the result does not depend on
the frequency $\o$ for reasons we have already discussed. 
   
In the case of slow field $y \ll 1$, but still 
$zy^2 \gg 1$ (strong field) we obtain
\be
\label{26}
\frac{ \Delta g_{wl} }
{ \Delta g^{(0)}_{wl} }=\frac{\Gamma(1/6)}{\pi\Gamma(5/6)}
\left(\frac{2\delta_1^2g_*^3}{9 C_0 \o^2}\right)^{1/3},
\ee
{\em i.e.}, the dependences both on the amplitude and frequency are 
different from the bulk case. 

In conclusion, we proposed
a random matrix theory describing influence of time dependent 
external field on the average magnetoresistance of open quantum
dots. This dependence can be recast in the form of the universal function
Eq.(\ref{22}) of one fitting parameter Eq.(\ref{18}) which can be fixed
by an independent experiment. 
The results can not be described by a simple replacement $g_\ast
\to g_\ast + \gamma_\phi$.
Finally, we mention that 
thermal fluctuations of the gate
potentials may induce the dephasing by virtue of the mechanism
considered here. However, the spectral density of such fluctuations
is model dependent which makes quantitative predictions hardly possible.

We acknowledge discussions with B.L. Altshuler,
V. Ambegaokar, and C.M. Marcus. 
Work was supported by
Cornell Center for Materials Research, funded under
NSF grant DMR-9632275  (MGV), and 
 A.P. Sloan and Packard foundations (ILA).

\end{multicols}

\end{document}